\documentclass{elsart}

\usepackage[square,comma]{natbib}
\usepackage{graphicx}
\usepackage{pxfonts}
\usepackage{lineno}

\usepackage{amssymb}

\journal{}

\begin{document}
\thispagestyle{empty}
\begin{Large}
\textbf{DEUTSCHES ELEKTRONEN-SYNCHROTRON}

\textbf{\large{Ein Forschungszentrum der
Helmholtz-Gemeinschaft}\\}
\end{Large}

DESY 10-108

July 2010

\begin{eqnarray}
\nonumber &&\cr \nonumber && \cr \nonumber &&\cr
\end{eqnarray}
\begin{eqnarray}
\nonumber
\end{eqnarray}
\begin{center}
\begin{Large}
\textbf{Scheme for generation of fully-coherent, TW power level hard X-ray pulses from baseline undulators at the European X-ray FEL}
\end{Large}
\begin{eqnarray}
\nonumber &&\cr \nonumber && \cr
\end{eqnarray}

\begin{large}
Gianluca Geloni,
\end{large}
\textsl{\\European XFEL GmbH, Hamburg}
\begin{large}

Vitali Kocharyan and Evgeni Saldin
\end{large}
\textsl{\\Deutsches Elektronen-Synchrotron DESY, Hamburg}
\begin{eqnarray}
\nonumber
\end{eqnarray}
\begin{eqnarray}
\nonumber
\end{eqnarray}
ISSN 0418-9833
\begin{eqnarray}
\nonumber
\end{eqnarray}
\begin{large}
\textbf{NOTKESTRASSE 85 - 22607 HAMBURG}
\end{large}
\end{center}
\clearpage
\newpage

\begin{frontmatter}



\title{Scheme for generation of fully-coherent, TW power level hard X-ray pulses from baseline undulators at the European X-ray FEL}


\author[XFEL]{Gianluca Geloni\thanksref{corr},}
\thanks[corr]{Corresponding Author. E-mail address: gianluca.geloni@xfel.eu}
\author[DESY]{Vitali Kocharyan}
\author[DESY]{and Evgeni Saldin}

\address[XFEL]{European XFEL GmbH, Hamburg, Germany}
\address[DESY]{Deutsches Elektronen-Synchrotron (DESY), Hamburg,
Germany}

\begin{abstract}
The most promising way to increase the output power of an X-ray FEL (XFEL) is by tapering the magnetic field of the undulator. Also, significant increase in power is achievable by starting the FEL process from a monochromatic seed rather than from noise. This report proposes to make use of a cascade self-seeding scheme with wake monochromators in a tunable-gap baseline undulator at the European XFEL to create a source capable of delivering coherent radiation  of unprecedented characteristics at hard X-ray wavelengths. Compared with SASE X-ray FEL parameters, the radiation from the new source has three truly unique aspects: complete longitudinal and transverse
coherence, and a peak brightness three orders of magnitude higher than what is presently available at LCLS. Additionally, the
new source will generate hard X-ray beam at extraordinary peak  (TW) and average (kW) power level. The proposed source can thus revolutionize fields like single biomolecule imaging, inelastic scattering and nuclear resonant scattering. The self-seeding scheme with the wake monochromator is extremely compact, and takes almost no cost and time to be implemented. The upgrade proposed in this paper could take place during the commissioning stage of the European XFEL, opening a vast new range of applications from the very beginning of operations.  We present feasibility study and examplifications for the SASE2 line of the European XFEL.
\end{abstract}

%
%
%
\end{frontmatter}



\section{\label{sec:intro} Introduction}

The LCLS, the first hard X-ray FEL in the world, was designed as a SASE amplifier, and produces very short and intense X-ray pulses \cite{LCLS1,LCLS2,DING}. Despite these results, some applications, including single biomolecule imaging, may require still higher photon flux \cite{tdr-2006} (Chapter 6, and references within).

Simulations indicate that if very short and very intense coherent X-ray pulses are available, a single scattering pattern could be recorded from a single protein molecule, before radiation damage starts to play a role and ultimately destroys the sample. Atomic resolution imaging is achievable  with pulse duration less than $5$ fs at a photon energy of $8$ keV \footnote{The optimal photon energy for diffraction imaging for pulses shorter
than 10 fs is 8 keV, \cite{tdr-2006}.}, and a number of photon per pulse greater than $10^{12}$. Besides full transverse coherence, full longitudinal coherence of the radiation pulse will be required in these experiments, in order to enable diffraction at large angles.

The most promising way to extend the output power in the low-charge (down to $0.025$ nC) mode of operation, is by tapering the magnetic field of the undulator, \cite{TAP1}-\cite{TAP4}. Further increase in power is achievable by starting the FEL process from a monochromatic seed, rather than from noise \cite{FAWL}. In this paper we propose to take advantage of a cascade self-seeding scheme with wake monochromators \cite{OURX,OURY2} in the baseline SASE2 undulator of the European XFEL \cite{tdr-2006} to create a source capable of delivering fully-coherent, $5$ fs-FWHM X-ray pulses with $2\cdot 10^{12}$ photons per pulse at $0.15$ nm wavelength. A self-seeding scheme with wake monochromators is inexpensive and compatible with the baseline design. The upgrade proposed in this paper could take place during the commissioning stage of the European XFEL. In this way, exciting biological problems would be open for investigation from the first day of operation of the European XFEL.

The proposed high-power mode of operation could be applied at higher bunch charge too. With a tunable-gap baseline undulator coupled to the self-seeding scheme, unprecedented  peak (up to $10^{37}$ ph/s/mm$^2$/mrad$^2$/$0.1\%$ BW) and average brightness (up to
$10^{28}$ ph/s/mm$^2$/mrad$^2$/$0.1\%$ BW) will be possible for charges of 0.25 nC, and many users could be served simultaneously by a single XFEL complex.
Using X-ray pulses from the proposed setup, techniques developed at third-generation radiation facilities - in particular those requiring high spectral purity - could be dramatically enhanced. Third generation synchrotron radiation sources in the hard X-ray range produce about $10^9$ photons per second with a meV bandwidth, which can be increased more than a million-fold at the European XFEL, by a combination of our self-seeding scheme and a 42-cells tunable-gap undulator. Techniques like inelastic X-ray scattering  (IXS) and nuclear resonance scattering (NMS), which are currently limited by the available photon flux in the desired bandwidth, can be
revolutionized by our self-seeding technique to enhance the longitudinal coherence.

Our work is organized in the following way. In the next Section we introduce the principle of the proposed technique. Based on our scheme, in the subsequent Section we discuss a way of obtaining a multi-user distribution system for an XFEL laboratory. In the two following sections we present a feasibility study for the short and long bunch mode of operation respectively. Finally we come to conclusions.

\section{\label{sec:method} Cascade self-seeding technique - key to generation of TW-level pulses in hard X-ray FEL}

\begin{figure}[tb]
\includegraphics[width=1.0\textwidth]{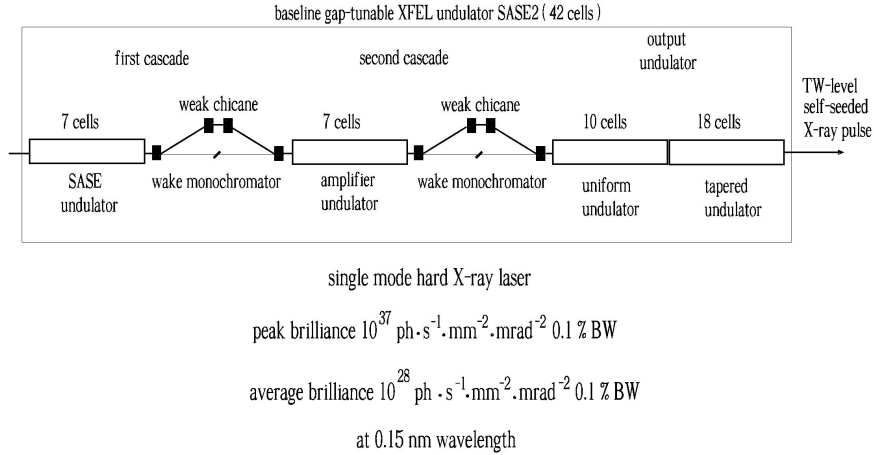}
\caption{Design of an undulator system for high power mode of operation. The method exploits a combination of a cascade self-seeding scheme with wake monochromators and an undulator tapering technique.} \label{csst1}
\end{figure}

\begin{figure}[tb]
\includegraphics[width=1.0\textwidth]{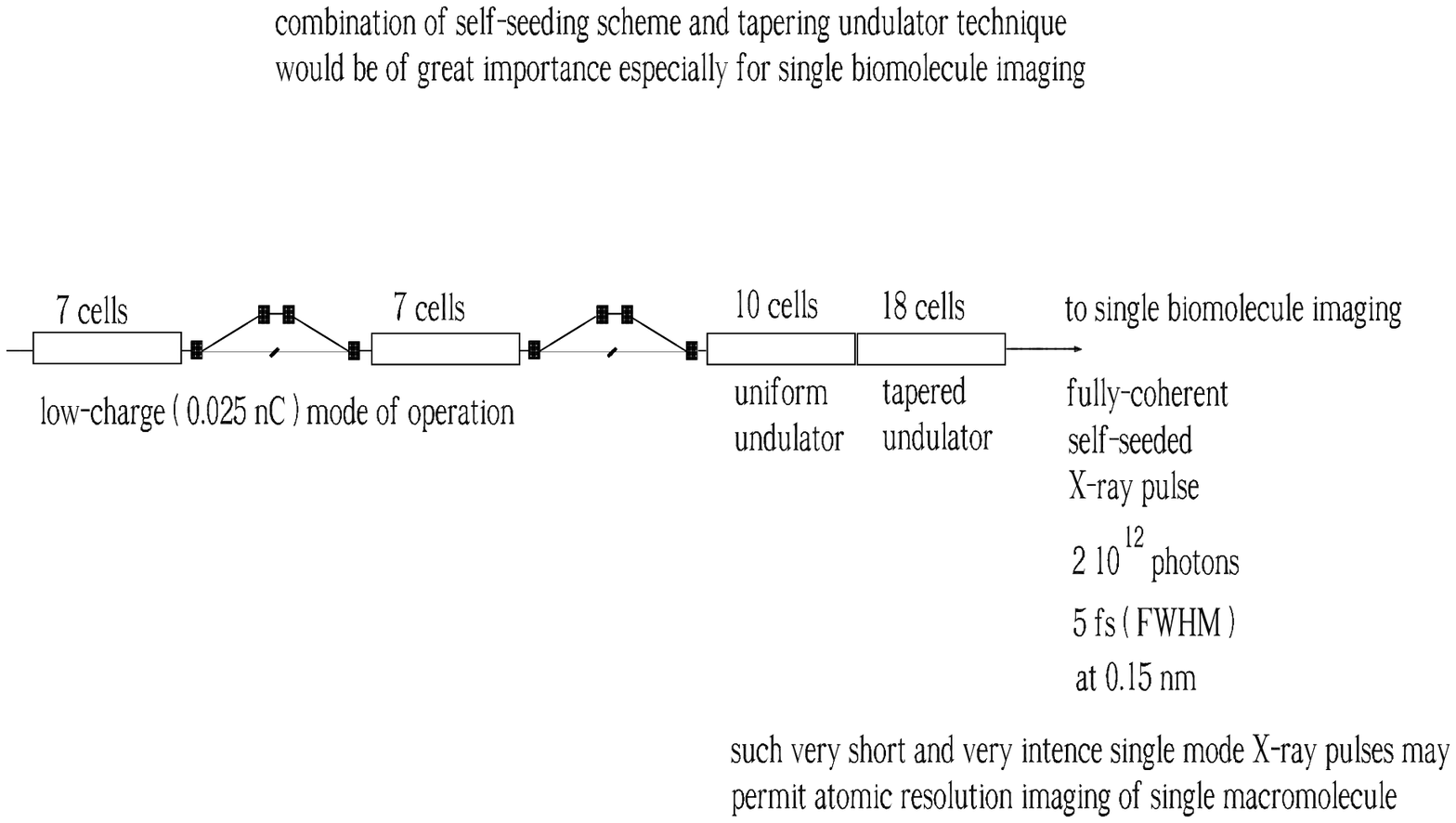}
\caption{The combination of high power and low charge mode of operation. This scheme holds a great promise as a source of X-ray radiation for such application as single biomolecule imaging.} \label{csst2}
\end{figure}

As mentioned before, the most promising way to increase the output power of the SASE radiation is by tapering the magnetic field of the undulator. Tapering consists in a slow reduction of the field strength of the undulator in order to preserve the resonance wavelength as the kinetic energy of the electrons decreases due to the FEL process. The strong radiation field produces a ponderomotive well which is deep enough to trap electrons. The radiation produced by these captured particles increases the depth of the ponderomotive well, so that electrons are effectively decelerated, leading to an increase of the radiated power. Tapering thus results in a much higher output power compared to the case of a uniform undulator. The undulator taper could be simply implemented as a step taper from one undulator segment to the next. The magnetic field tapering is provided by changing the undulator gap. A further increase in power is achievable by starting the FEL process from a monochromatic seed, rather than from noise. The reason is the higher degree of coherence of the radiation in the seed case, which involves, with tapering, a large portion of the bunch in the energy-wavelength synchronism. Here we propose a scheme for generation of TW-level X-ray pulses in tapered XFELs with the use of the cascade self-seeding technique for highly monochromatic seed generation. In this way, considering for example the SASE2 undulator at the European XFEL, the output power of the XFEL radiation could be increased from the baseline value of $20-30$ GW to about $400-500$ GW.

Fig. \ref{csst1} shows the design principle of our setup for high power mode of operation. Two wake-monochromator cascades, identical to those considered in \cite{OURY2}, are followed by an output undulator consisting of two sections.  The first section is composed by a $60$ m long (10 cells) uniform undulator, the second
section by a $108$ m long (18 cells) tapered undulator.  A nearly Fourier limited radiation pulse is produced in the first two cascades, which is then exponentially amplified passing through the first (uniform) section of the output undulator. This section is long enough to reach saturation, which yields about $20$ GW power. Finally, in the second section, the monochromatic FEL output is enhanced up to $400-500$ GW, by means of a $2 \%$ tapering of the undulator parameter over the last $18$ cells after saturation. Such result can be reached at the baseline tunable-gap undulator SASE2 (42 cells), with the help of our self-seeding technique based on the use of single crystal monochromators.

The proposed high-power mode of operation could be applied at the operating charge of $0.25$ nC. The output characteristics in this case would be extraordinary for a SASE XFEL. The peak brightness of such source is about three orders of magnitude higher than what is currently available at LCLS, yielding up to $10^{37}$ ph/s/mm$^2$/mrad$^2$/$0.1\%$ BW, while the average brightness may be up to $10^{28}$ ph/s/mm$^2$/mrad$^2$/$0.1\%$ BW , i.e. 7 orders of magnitude higher compared to present day synchrotron radiation sources.

We can also apply this scheme to low-charge mode (0.025 nC) to obtain fully-coherent TW-level X-ray pulses with duration
less than 5 f (see Fig. \ref{csst2}). Theoretical studies and simulations \cite{tdr-2006} (Chapter 6 and references therein) predict that such pulses may allow structural studies on large biomolecules before the radiation damage destroys the sample.

\section{\label{sec:multi} Cascade self-seeding  technique - key to obtain multi-user distribution system for XFEL laboratory}

\begin{figure}[tb]
\includegraphics[width=1.0\textwidth]{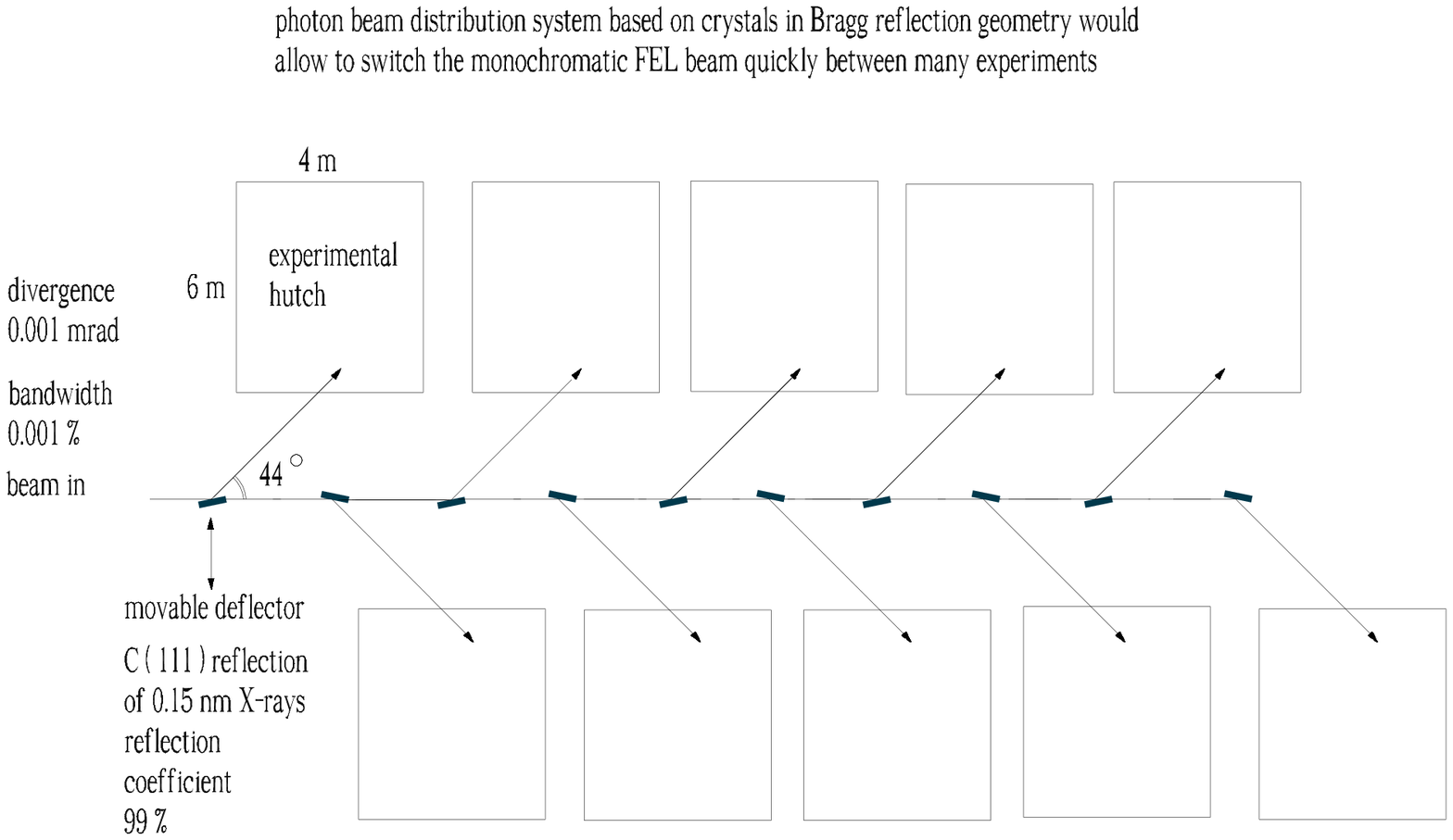}
\caption{Proposed hard X-ray FEL undulator beam line. A photon beam distribution system based on movable crystals can provide an efficient way to obtain a multi-user facility.} \label{csst3}
\end{figure}

\begin{figure}[tb]
\includegraphics[width=1.0\textwidth]{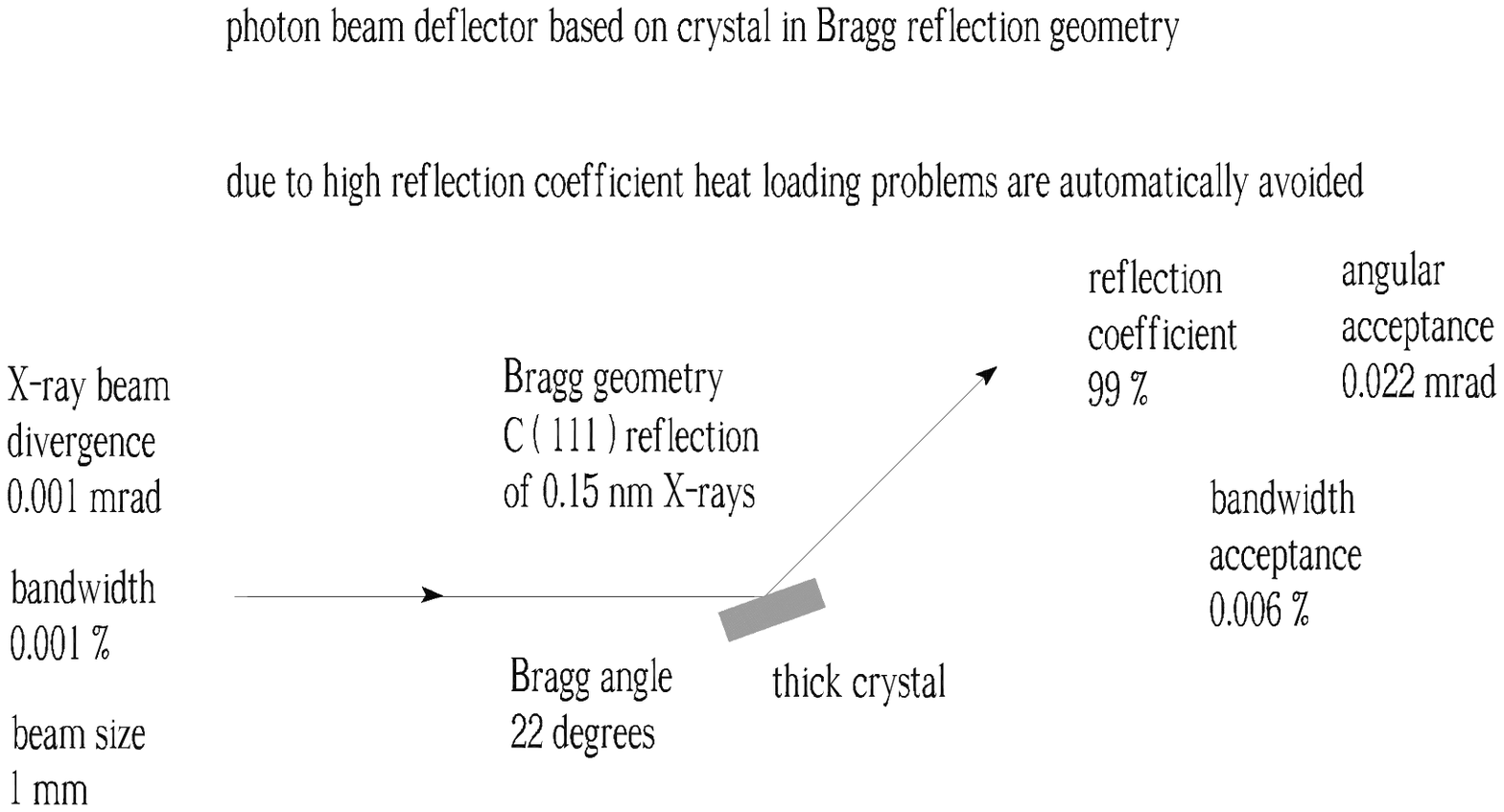}
\caption{Concept of the photon beam deflector based on the use of a crystal in Bragg reflection geometry.} \label{csst4}
\end{figure}
As discussed in the previous Sections, in this paper we propose a method allowing to increase the peak and average brilliance by three orders of magnitude compared to baseline XFEL specifications.
The combination of very high peak-power and very narrow bandwidth, based on a cascade self-seeding scheme for monochromatization of the X-rays, and on an undulator tapering technique will open a vast new range of applications.

An advantage of the proposed scheme is the possibility to obtain a multi-user facility. The typical layout of a SASE FEL is a linear arrangement, in which injector, accelerator, bunch compressors
and undulators are nearly collinear, so that the electron beam does not change its direction along the setup. However, it is obviously desirable that an X-ray FEL laboratory could serve tens of
experimental stations, which should operate independently according to the  needs of the user community. In this section we describe a photon beam distribution system, sketched in Fig. \ref{csst3}, which may allow to switch the FEL beam quickly among many experiments in order to make an effective use of the facility.

The technical approach adopted in this variation of the XFEL laboratory design makes use of movable thick diamond crystals
in Bragg reflection geometry. For the C(111) reflection, the angular
acceptance of the crystal deflector is of the order of
$20$ microradians and the spectral bandwidth is about $6 \cdot 10^{-5}$. As a result, the angular and frequency acceptance of the deflector is much wider compared to the photon beam divergence, which is of the order of a microradian, and bandwidth of the order of 0.001 \%, and it is therefore possible  to deflect the full photon beam without perturbations.
About $99 \%$ of the peak reflectivity can be achieved for wavelengths around 0.15 nm. It should be noted that the deflection process happens only once during the pass of a given photon pulse  through the deflector unit. This is because single crystal provides a sufficiently large deflection angle (of the order of $\pi/4$), so that the problem of absorption of the radiation in the distribution system does not exist.

A schematic of a movable deflector is shown in Fig. \ref{csst4}.

\section{\label{sec:feasis} Feasibility study for short bunch mode of operation}

Following the previous introduction of the proposed methods we report on a feasibility study of the single-bunch self-seeding scheme combined with tapering. This feasibility study is performed with the
help of the FEL code GENESIS 1.3 \cite{GENE} running on a parallel
machine. In this section we will present the feasibility
study for the short-pulse mode of operation ($1~\mu$m rms) while, later on, we will cover the long-pulse mode of operation ($10~\mu$m rms). Parameters used in the simulations for the short pulse mode of operation are presented in Table \ref{tt1}. For the long pulse mode of operations Table \ref{tt1} is still valid, except for a ten times larger charge ($0.25$ nC) and a ten times longer rms bunch length.  We present a statistical analysis consisting of $100$ runs for the short mode of operation and of $50$ runs for the long pulse mode of operation.

\begin{table}
\caption{Parameters for the short pulse mode of operation used in
this paper.}

\begin{small}\begin{tabular}{ l c c}
\hline & ~ Units &  ~ \\ \hline
Undulator period      & mm                  & 48     \\
K parameter (rms)     & -                   & 2.516  \\
Wavelength            & nm                  & 0.15   \\
Energy                & GeV                 & 17.5   \\
Charge                & nC                  & 0.025 \\
Bunch length (rms)    & $\mu$m              & 1.0    \\
Normalized emittance  & mm~mrad             & 0.4    \\
Energy spread         & MeV                 & 1.5    \\
\hline
\end{tabular}\end{small}
\label{tt1}
\end{table}

\begin{figure}[tb]
\includegraphics[width=1.0\textwidth]{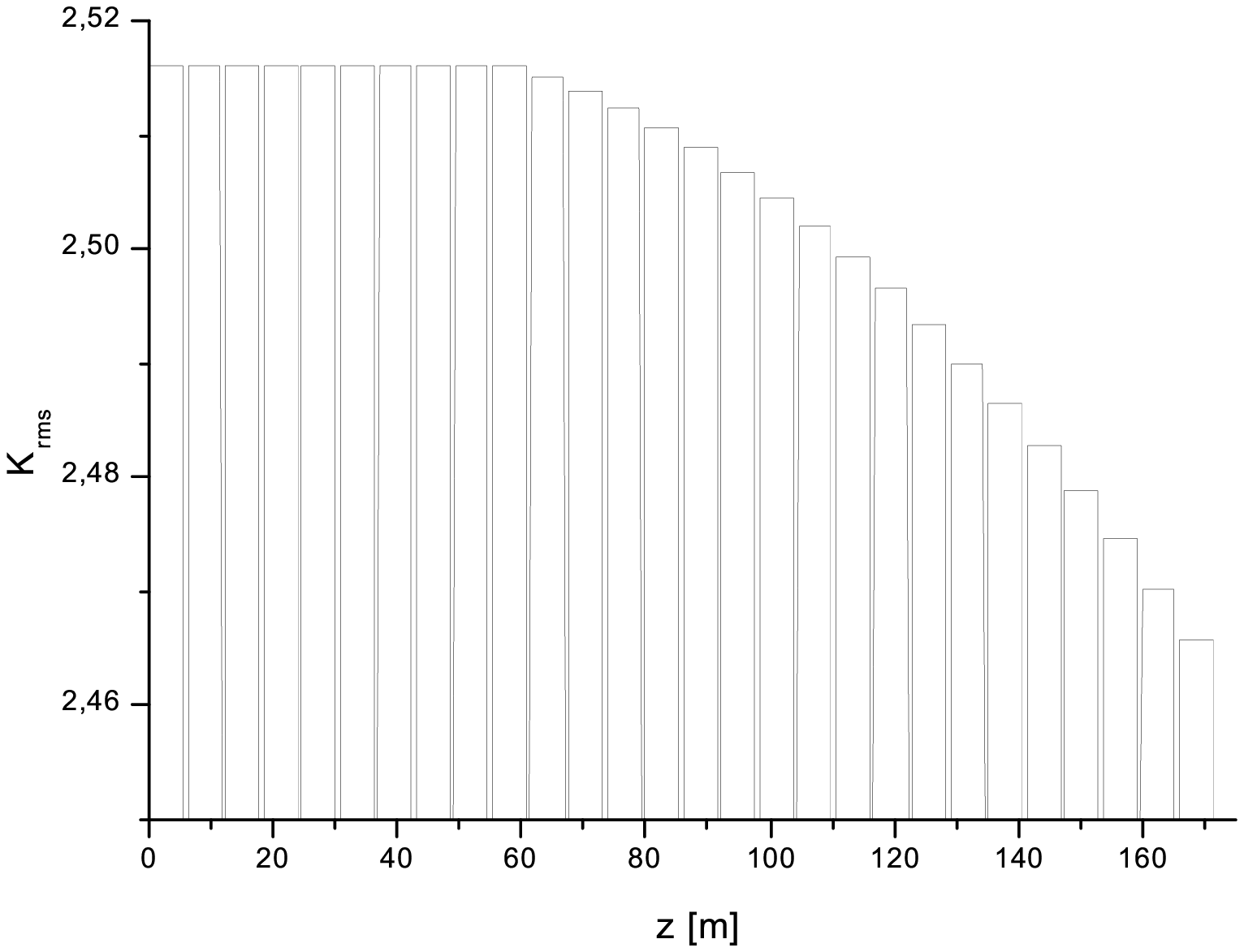}
\caption{Taper configuration for the high power mode of
operation at 0.15 nm.} \label{Taplaw}
\end{figure}

The value of the rms $K$ parameter refers to the untapered sections. The optimal \footnote{Note that additional tapering should be considered to keep undulators tuned in the presence of
energy loss from spontaneous synchrotron radiation. In Fig. \ref{Taplaw} we present a tapering configuration which is to be considered as an addition to this energy-loss compensation
tapering.} tapering configuration is presented, instead, in Fig. \ref{Taplaw}. We used the same tapering law for both short bunch and long bunch. Such law has been determined phenomenologically, with the help of numerical experiments. The effective value of K is kept constant through each  undulator segment. In other words, tapering consists of a stepwise change of K from segment to segment (segment
tapering).

\begin{figure}[tb]
\includegraphics[width=1.0\textwidth]{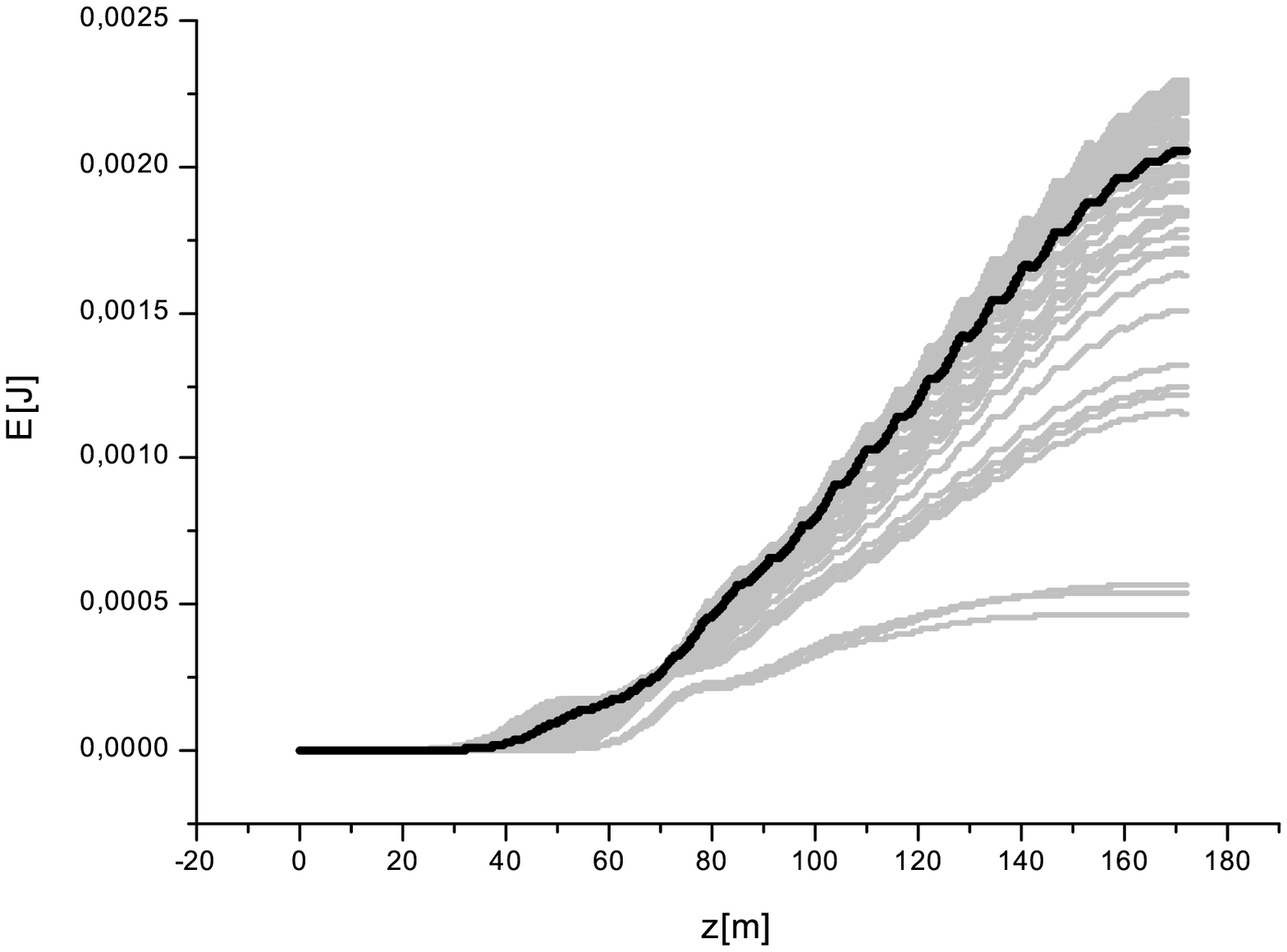}
\caption{Short bunch mode of operation. Energy in the X-ray radiation pulse versus the length of  the output undulator in the case of undulator tapering. Grey lines refer to single shot realizations, the black line refers to the average over a hundred realizations.} \label{Energy_S}
\end{figure}

\begin{figure}[tb]
\includegraphics[width=1.0\textwidth]{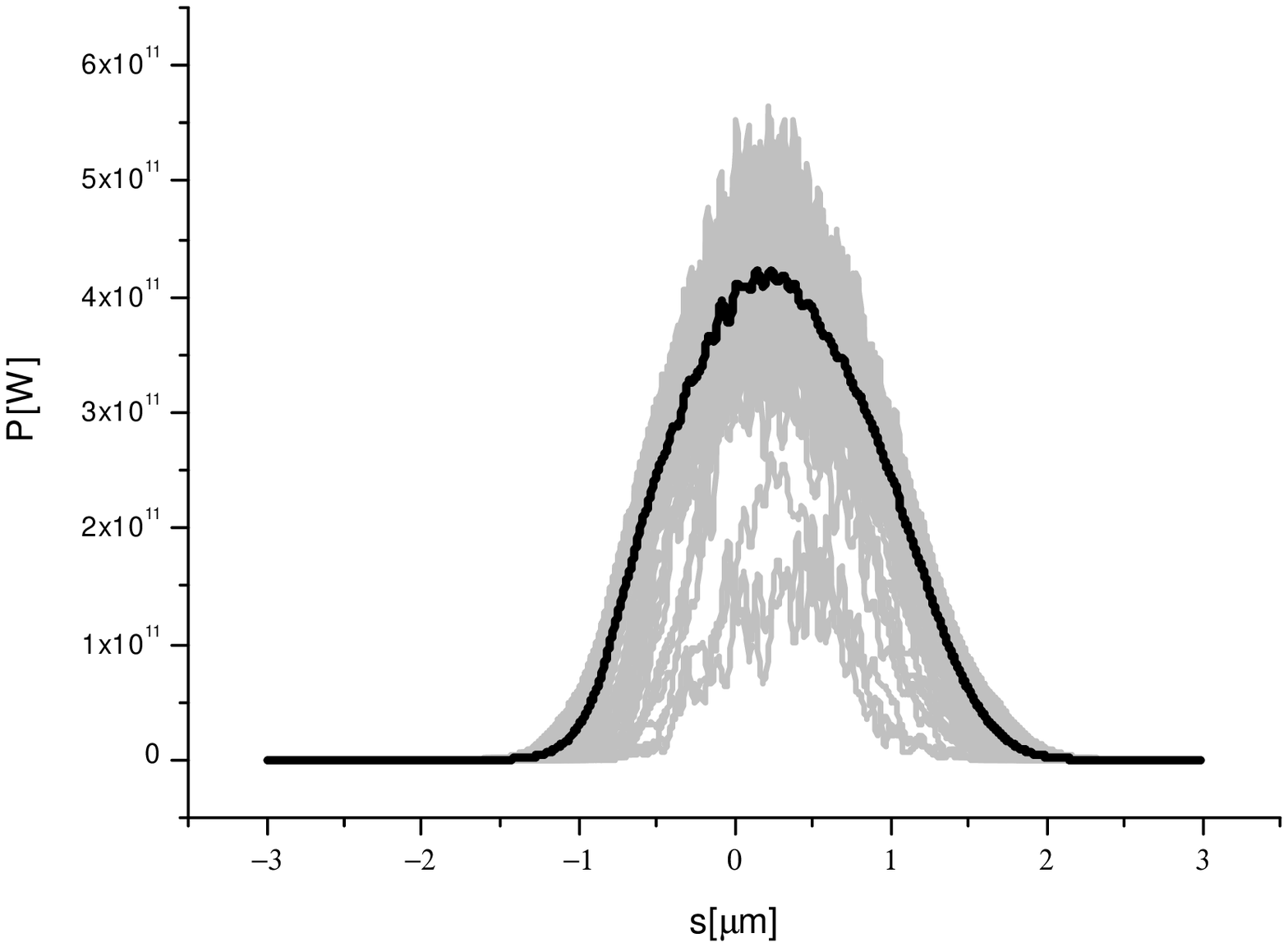}
\caption{Short bunch mode of operation. Power distribution of the X-ray radiation pulse in the case of undulator tapering. Grey lines refer to single shot realizations, the black line refers to the average over a hundred realizations.} \label{Pow_S}
\end{figure}

\begin{figure}[tb]
\includegraphics[width=1.0\textwidth]{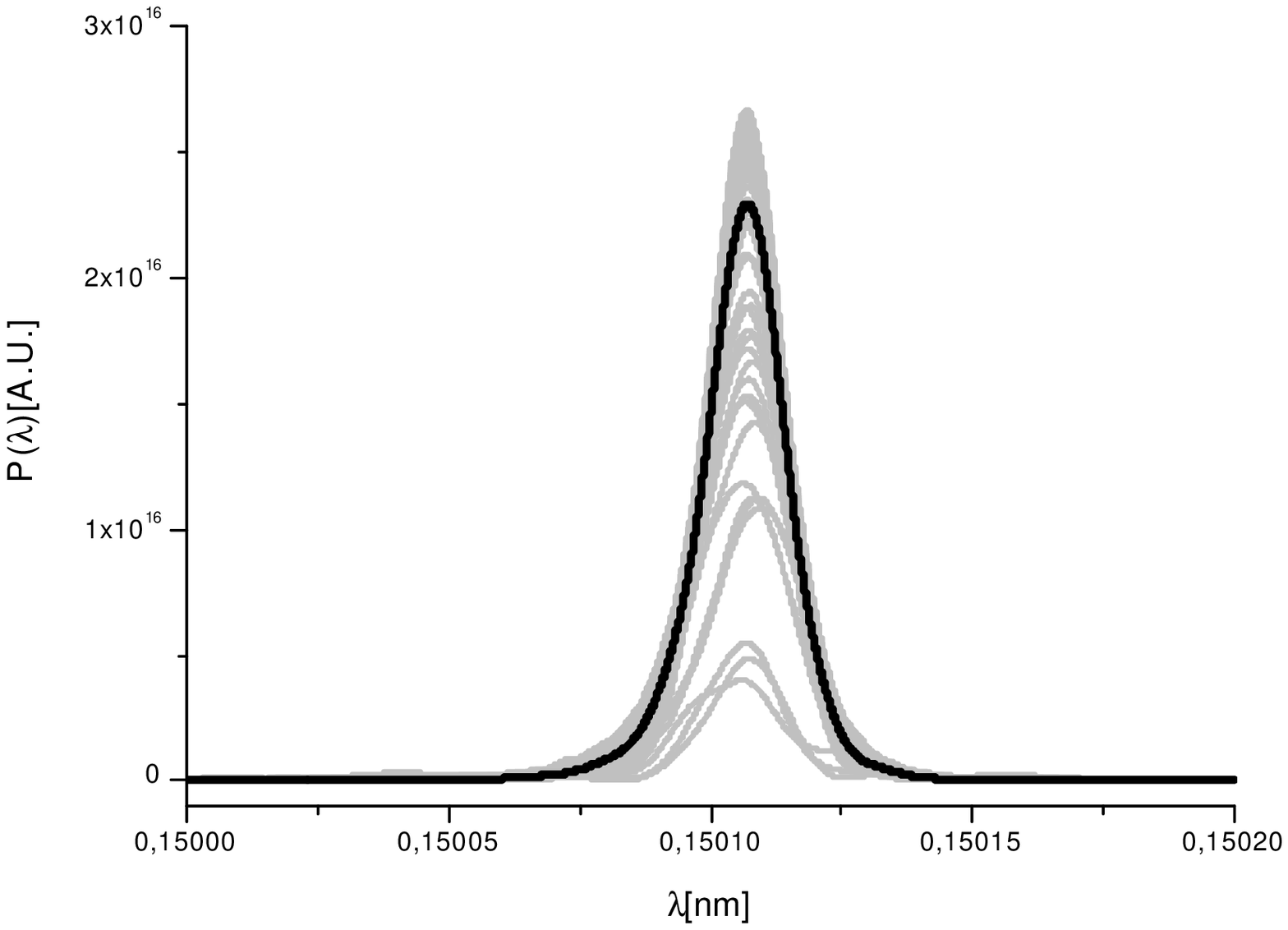}
\caption{Short bunch mode of operation. Spectrum of the X-ray radiation pulse in the case of undulator tapering. Grey lines refer to single shot realizations, the black line refers to the average over a hundred realizations.} \label{Spec_S}
\end{figure}

\begin{figure}[tb]
\includegraphics[width=1.0\textwidth]{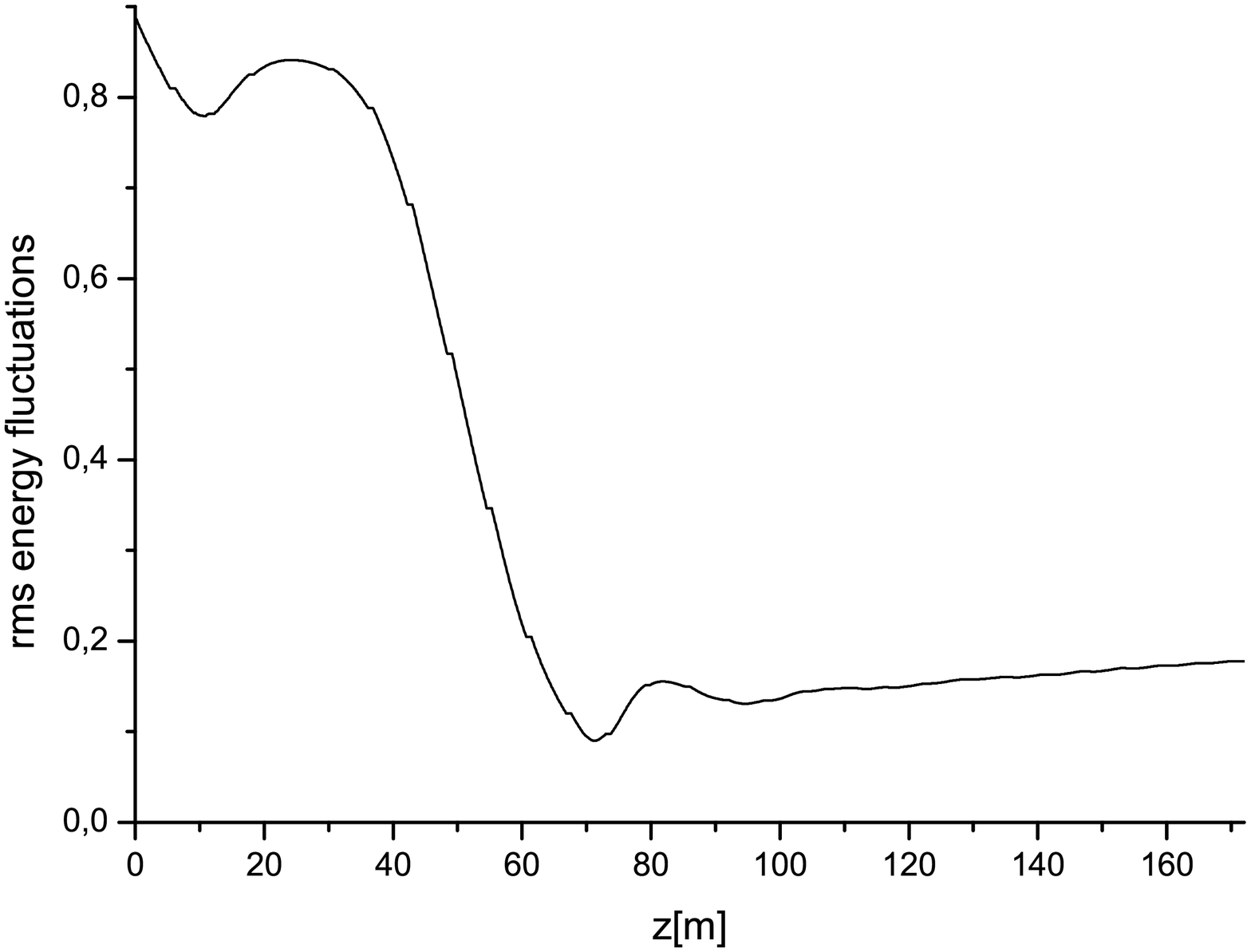}
\caption{Short bunch mode of operation. rms energy deviation from the average as a function of the distance inside the output undulator, composed of 10 uniform plus 18 tapered cells.} \label{SVar}
\end{figure}

After the first seven cells (42 m) the electron bunch is sent through the weak chicane, while radiation is filtered through a single diamond crystal, as shown in Fig. \ref{csst1}. We use the C(400) Bragg reflection and we assume that the crystal has a thickness of $0.1$ mm. Radiation and electron bunch are then recombined, and sent through another identical monochromatization cascade. The advantage of using two cascades is related to the high contrast between seeded and SASE signal. The two-cascade self-seeding technique has been described in detail in \cite{OURY2}. Following the second cascade, the seeded electron bunch enters an uniform undulator with ten cells (60 m) followed by an 18 cells tapered section (108 m), for a total undulator length of 252 m (42 cells).  Fig. \ref{Energy_S} shows the increase of energy in the radiation pulse as a function of the position inside the undulator, while Fig. \ref{Pow_S} and Fig. \ref{Spec_S} respectively show the output power and spectrum. A two-mJ, ultra-short (5 fs FWHM)  pulse with a bandwidth of about $1.2 \cdot 10^{-4}$ is produced. The rms energy deviation from the average as a function of the distance inside the output undulator is shown in Fig. \ref{SVar}. Fluctuations at the output of the device are about $20\%$.

\section{\label{sec:feasil} Feasibility study for long bunch mode of operation}

Similarly as for the case of the short bunch mode of operation, we conducted a feasibility study also for the long bunch mode of operation. Parameters are as in Table \ref{tt1}, the only difference being a larger bunch charge ($0.1$ nC), and a longer rms bunch length ($10 ~\mu$m).  As said above, we used the same tapering law as for the short bunch mode of operation, presented in Fig. \ref{Taplaw}. The setup used is identical as in the short bunch mode of operation.

\begin{figure}[tb]
\includegraphics[width=1.0\textwidth]{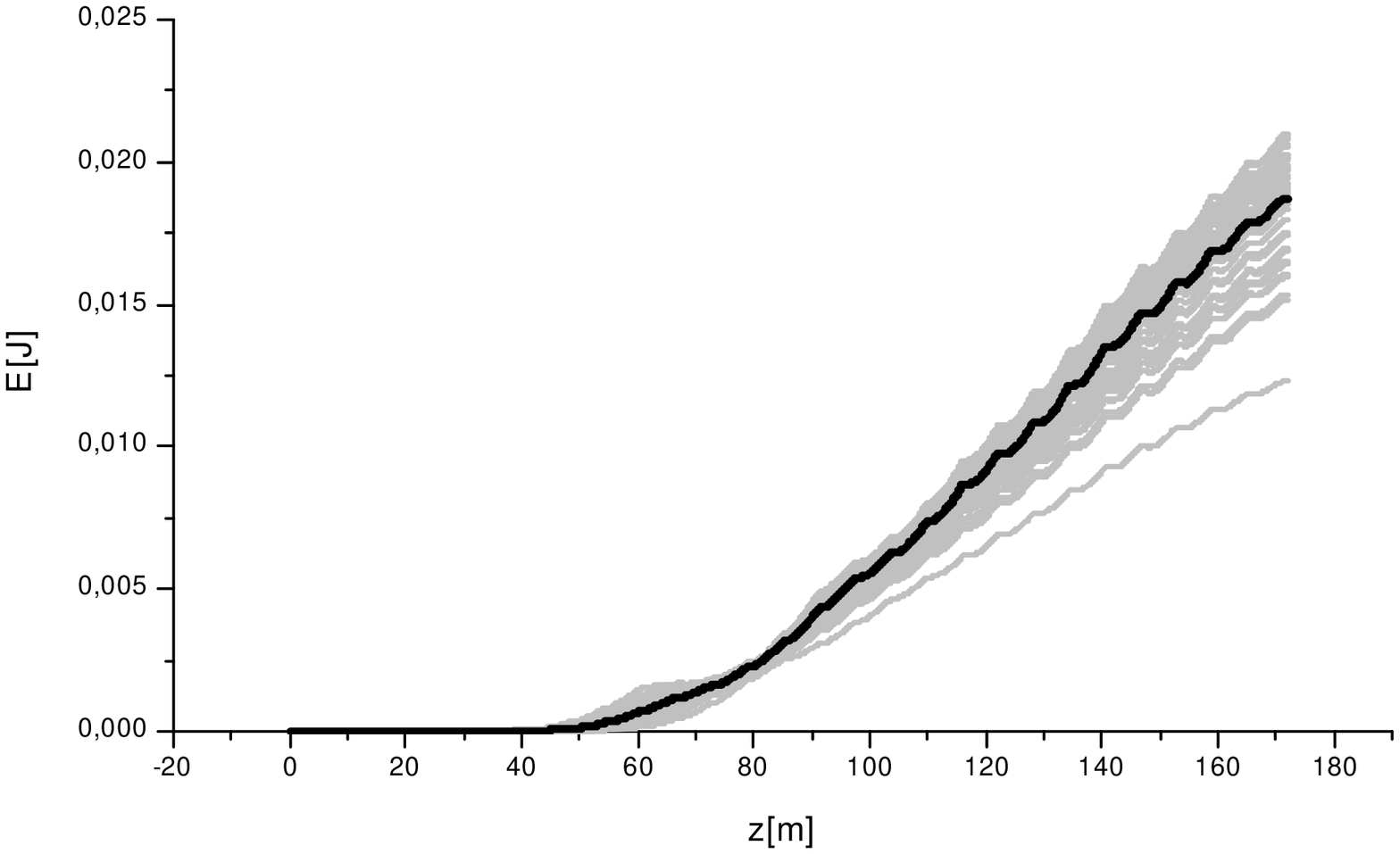}
\caption{Long bunch mode of operation. Energy in the X-ray radiation pulse versus the length of  the output undulator in the case of undulator tapering. Grey lines refer to single shot realizations, the black line refers to the average over fifty realizations.} \label{Energy_L}
\end{figure}

\begin{figure}[tb]
\includegraphics[width=1.0\textwidth]{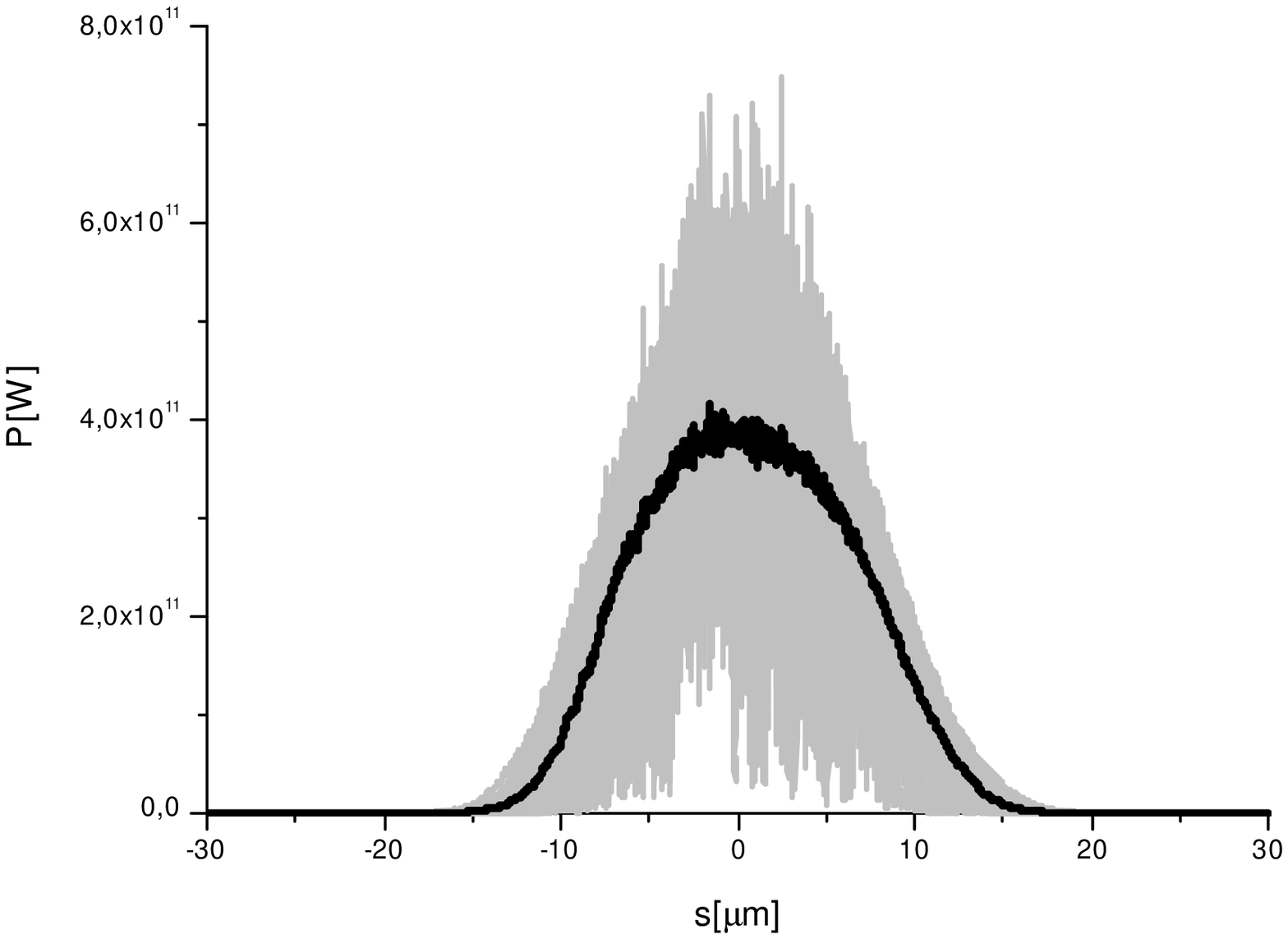}
\caption{Long bunch mode of operation. Power distribution of the X-ray radiation pulse in the case of undulator tapering. Grey lines refer to single shot realizations, the black line refers to the average over fifty realizations.} \label{Pow_L}
\end{figure}

\begin{figure}[tb]
\includegraphics[width=1.0\textwidth]{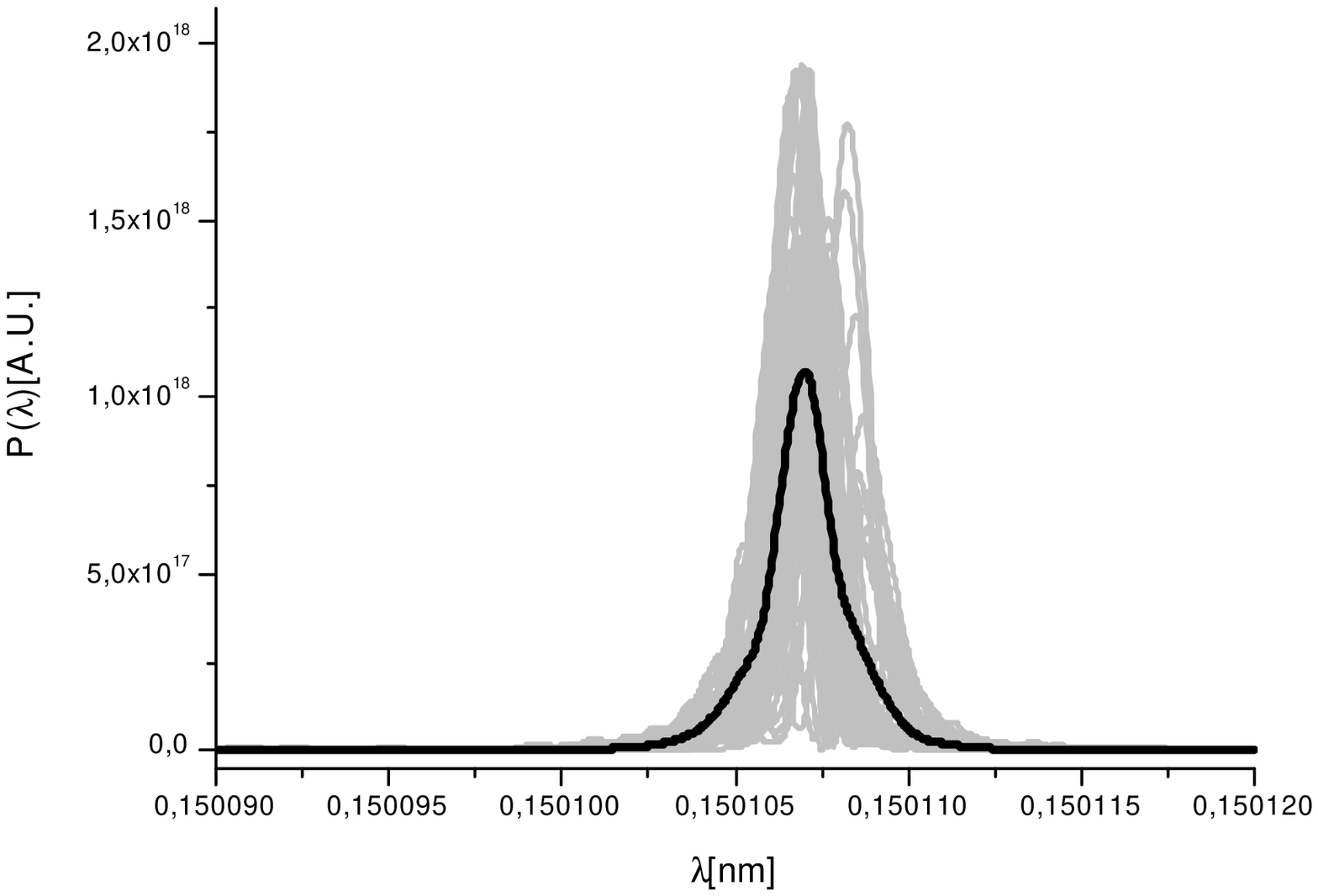}
\caption{Long bunch mode of operation. Spectrum of the X-ray radiation pulse in the case of undulator tapering. Grey lines refer to single shot realizations, the black line refers to the average over fifty realizations.} \label{Spec_L}
\end{figure}

\begin{figure}[tb]
\includegraphics[width=1.0\textwidth]{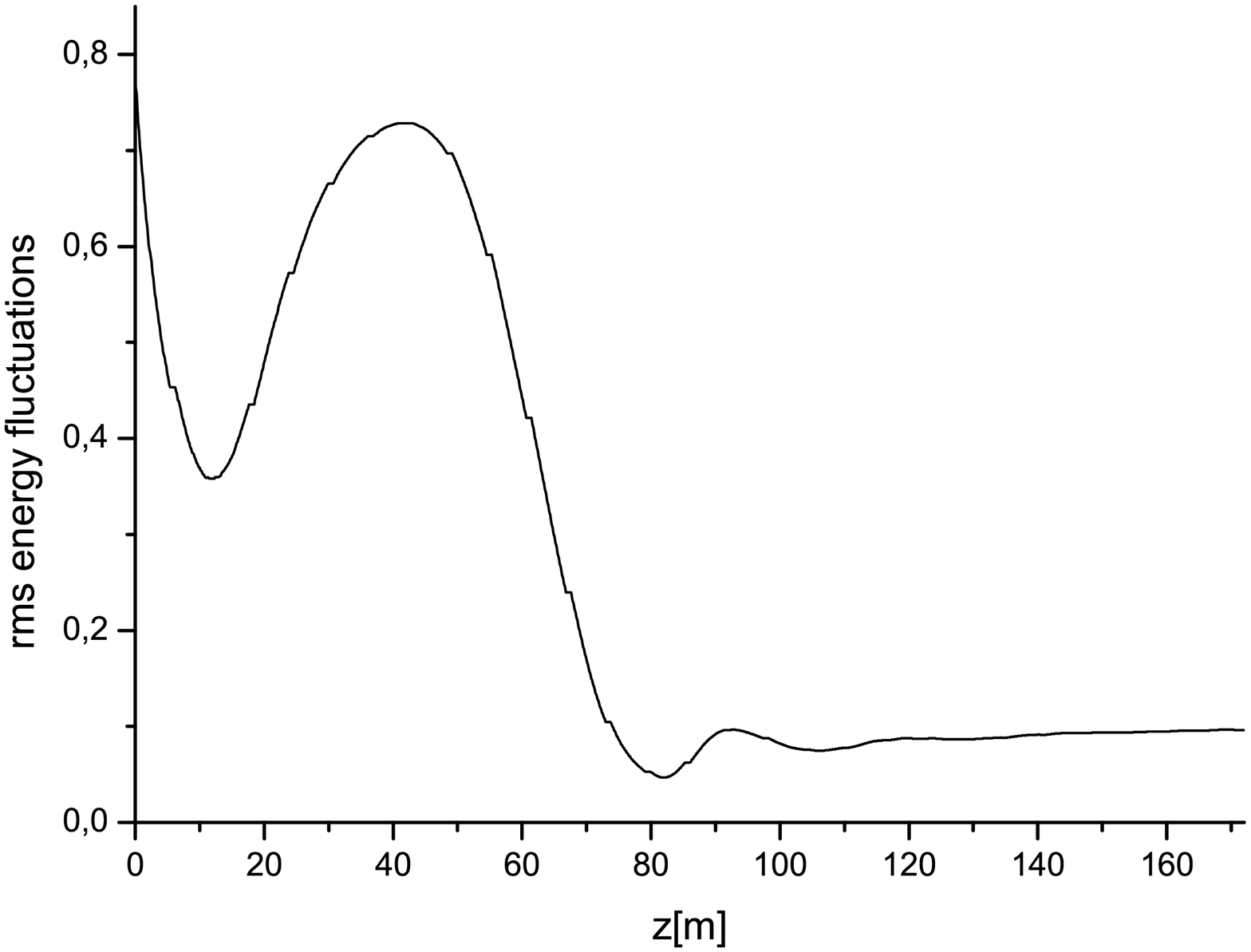}
\caption{Long bunch mode of operation. rms energy deviation from the average as a function of the distance inside the output undulator, composed of 10 uniform plus 18 tapered cells.} \label{LVar}
\end{figure}

Fig. \ref{Energy_L} shows the increase of energy in the radiation pulse as a function of the position inside the undulator, while Fig. \ref{Pow_L} and Fig. \ref{Spec_L} respectively show the output power and spectrum. A 20 mJ, TW-power level pulse with a bandwidth of order $10^{-5}$ is produced.

\begin{figure}[tb]
\includegraphics[width=1.0\textwidth]{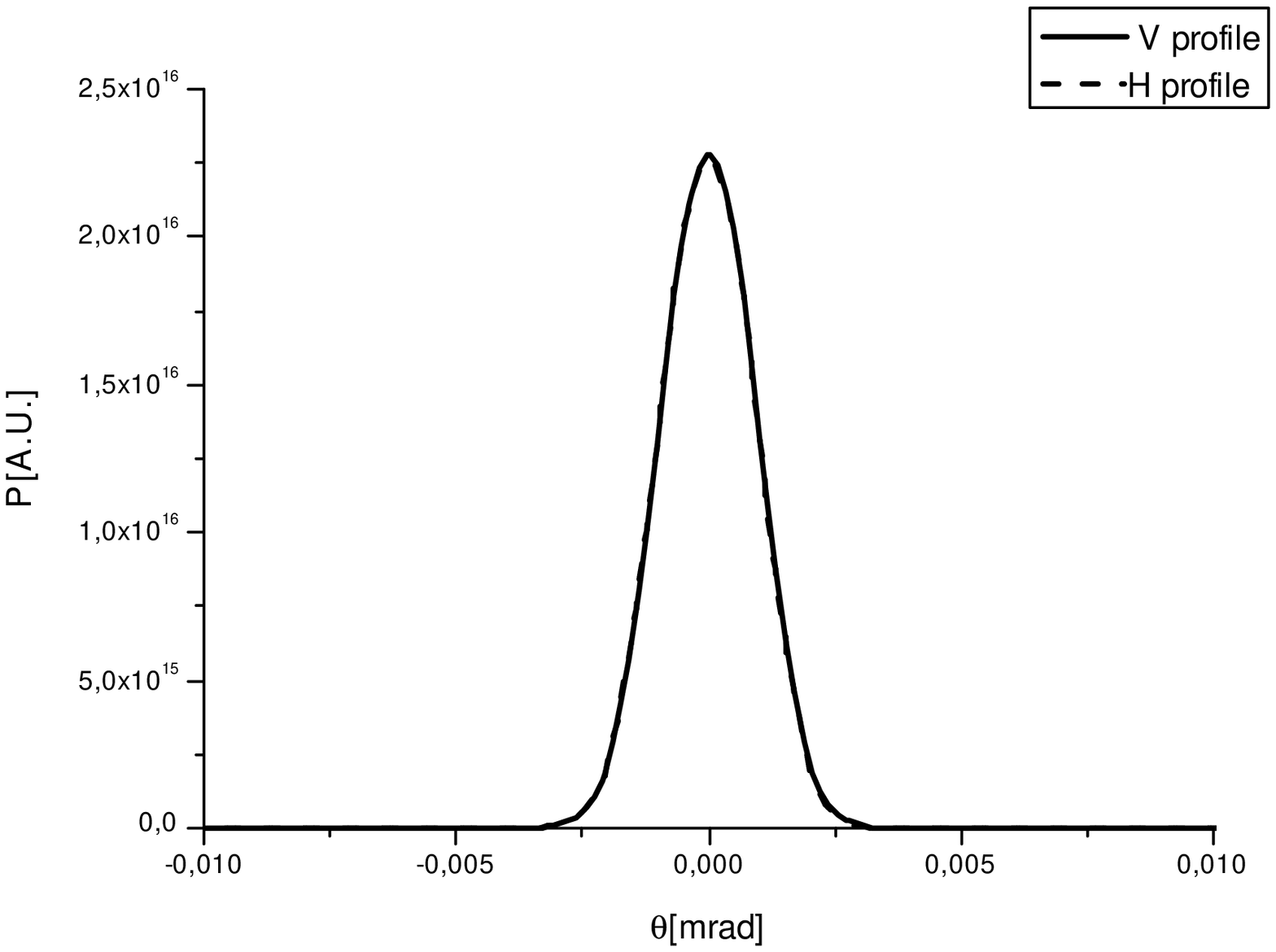}
\caption{Long pulse mode of operation. Angular distribution of X-ray radiation pulse energy at saturation without tapering. Here the output undulator is 12-cells long.} \label{Pow_notap_far_L}
\end{figure}

\begin{figure}[tb]
\includegraphics[width=1.0\textwidth]{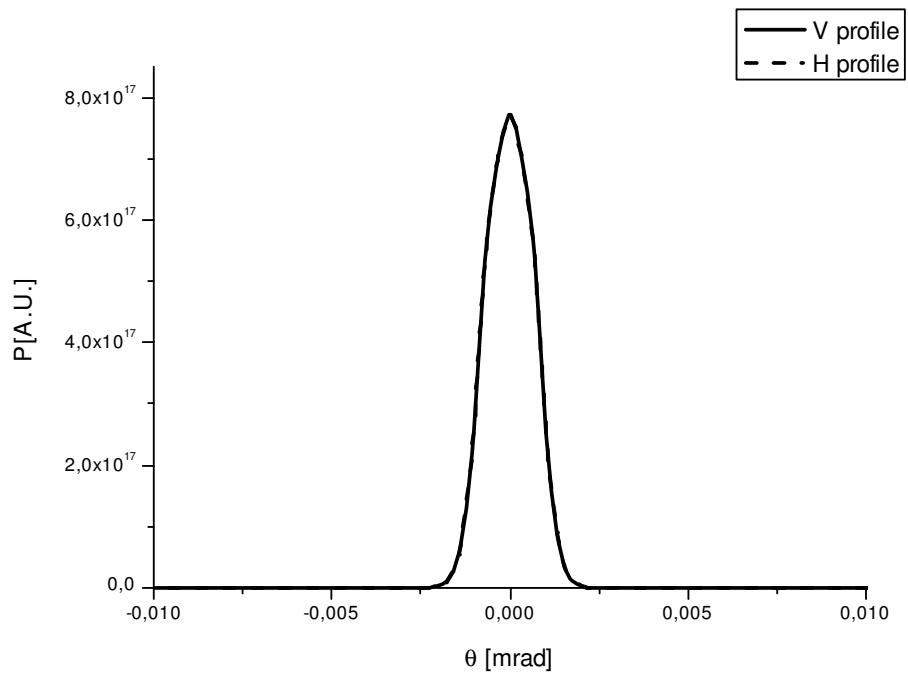}
\caption{Long pulse mode of operation. Angular distribution of X-ray radiation pulse energy in the case of undulator tapering. Here the output undulator is 28 cells-long.} \label{Pow_tap_far_L}
\end{figure}

\begin{figure}[tb]
\includegraphics[width=1.0\textwidth]{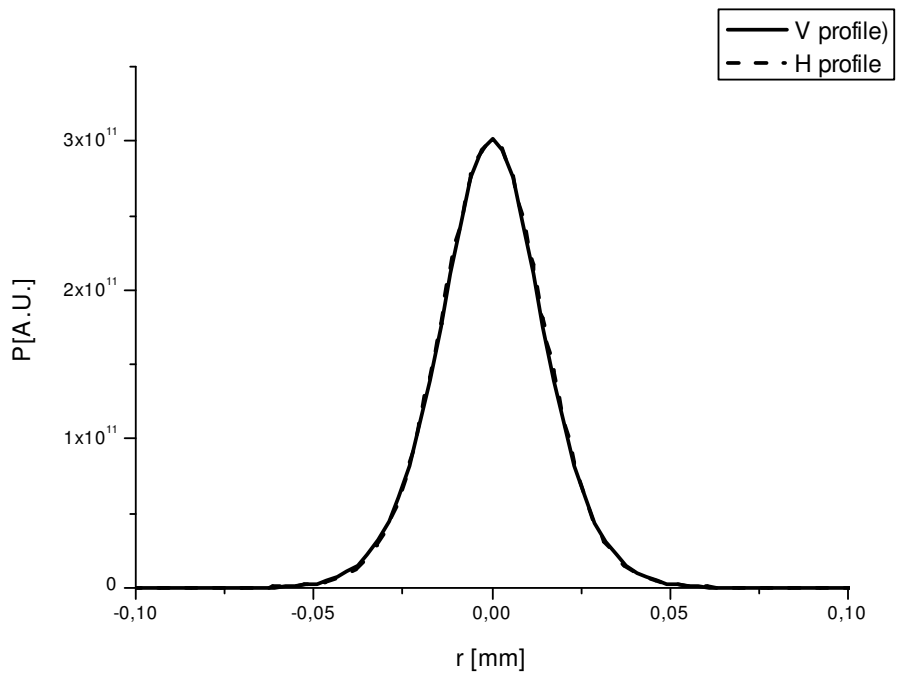}
\caption{Long bunch mode of operation. X-ray radiation pulse energy distribution per unit surface versus the radius at saturation without tapering. Here the output undulator is 12-cells long.} \label{Pow_notap_near_L}
\end{figure}

\begin{figure}[tb]
\includegraphics[width=1.0\textwidth]{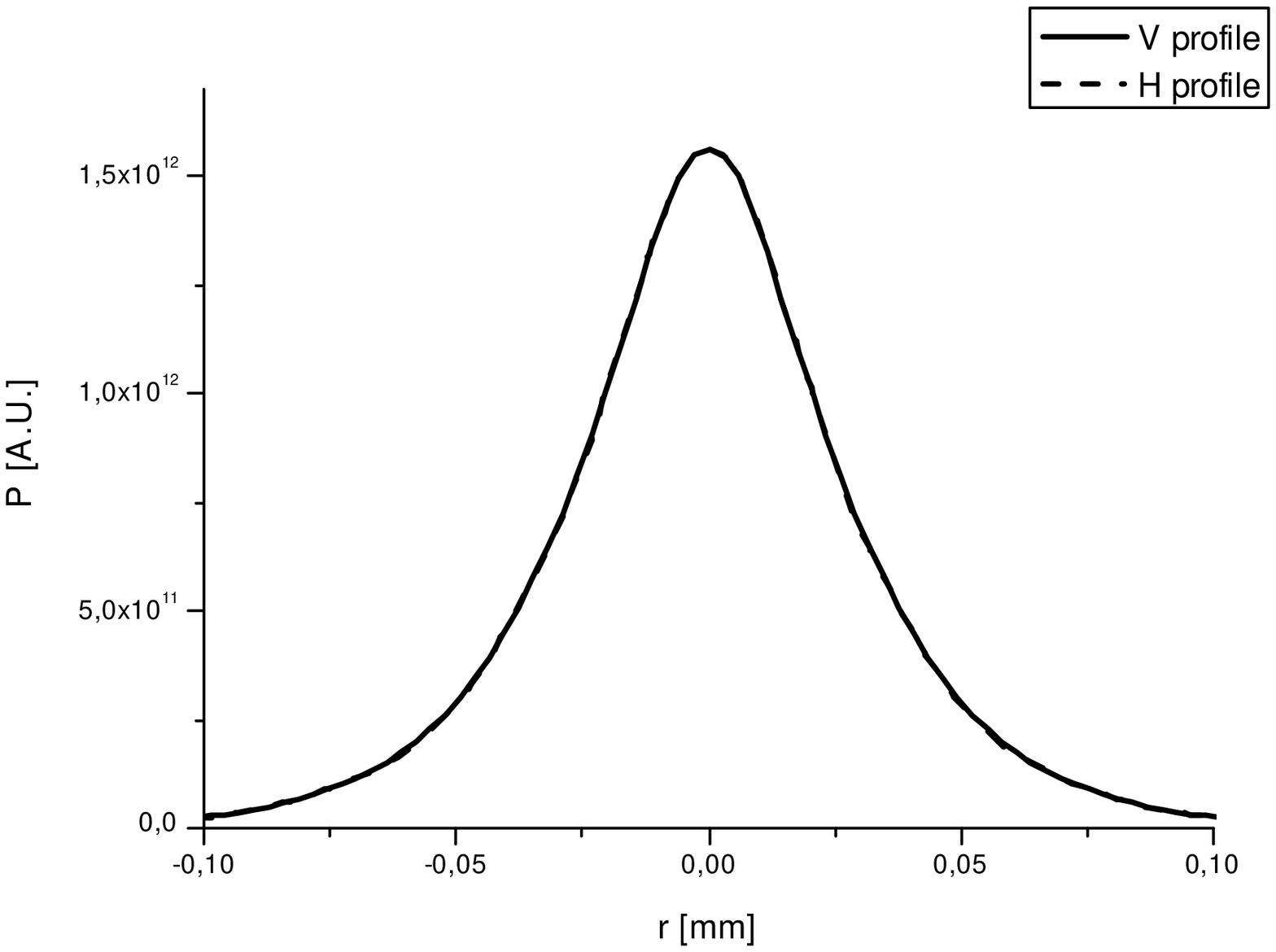}
\caption{Long pulse mode of operation. X-ray radiation pulse energy per unit surface distribution versus the radius at the output undulator exit in the case of undulator tapering. Here the output undulator is 28-cells long.} \label{Pow_tap_near_L}
\end{figure}

Since we performed 3-D simulations, we can calculate the distribution of the radiation field in the near and in the far diffraction zone as well. In particular, it is interesting to compare the angular distribution of the X-ray radiation pulse energy, and the transverse distribution of the radiation pulse energy per unit surface versus the radius, when the XFEL operates at saturation, both in the cases with and without tapering. Fig. \ref{Pow_notap_far_L} and Fig. \ref{Pow_tap_far_L} show the angular distributions respectively without (12-cells) and with (28-cells) tapering. Fig. \ref{Pow_notap_near_L} and Fig. \ref{Pow_tap_near_L} show, instead,  the energy distribution per unit surface versus the radius at saturation, in the cases without and with tapering respectively. The radiation spot size at the exit of the tapered undulator is
larger than that at saturation without tapering, and the radiation field expands out of the electron beam (see Fig. \ref{Pow_notap_near_L} and Fig. \ref{Pow_tap_near_L}). As a
result, the width of the angular distribution becomes narrower (see Fig. \ref{Pow_notap_far_L} and Fig. \ref{Pow_tap_far_L}). The rms energy deviation from the average as a function of the distance inside the output undulator is shown in Fig. \ref{LVar}. Fluctuations at the output of the device are about $10\%$.

\begin{figure}[tb]
\includegraphics[width=1.0\textwidth]{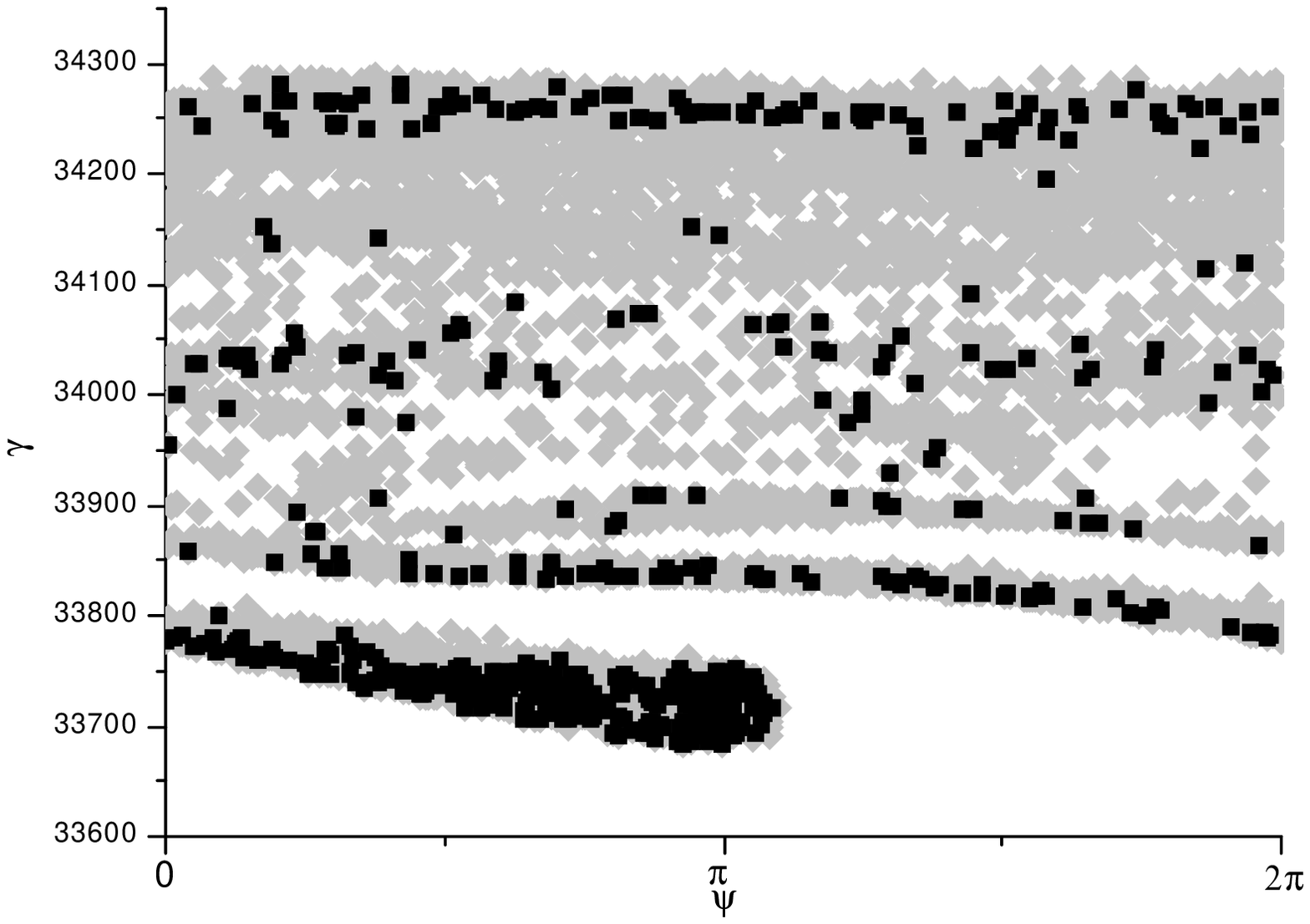}
\caption{Phase space distribution of the particles in the middle slice of the electron bunch at the undulator exit in the case of undulator tapering according to Fig. \ref{Taplaw}. The black spots represent macroparticles near the electron beam axis. The grey spots represent all macroparticles in the middle slice.} \label{Phspace_L}
\end{figure}

\begin{figure}[tb]
\includegraphics[width=1.0\textwidth]{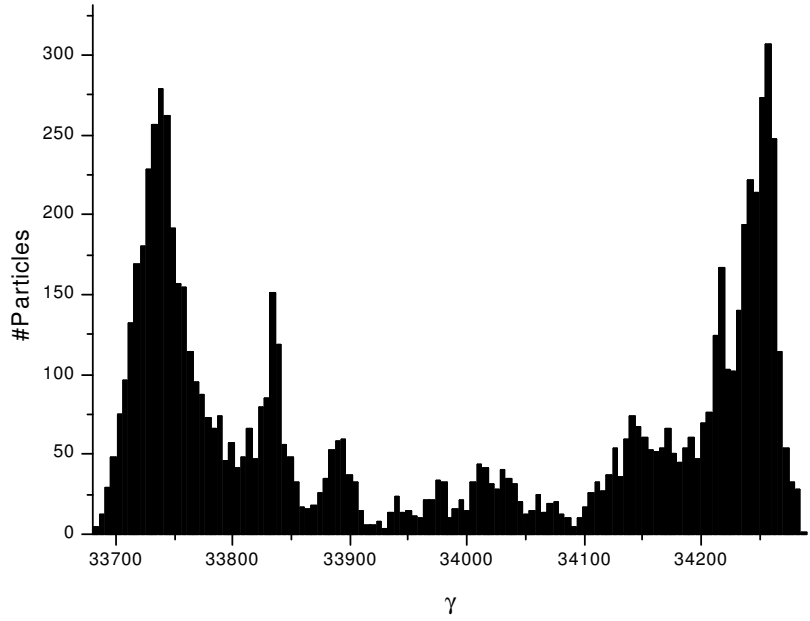}
\caption{Energy distribution of the particles in the middle
slice of the electron bunch at the undulator exit in the case of undulator tapering according to Fig. \ref{Taplaw}.} \label{Distr_middle_L}
\end{figure}

\begin{figure}[tb]
\includegraphics[width=1.0\textwidth]{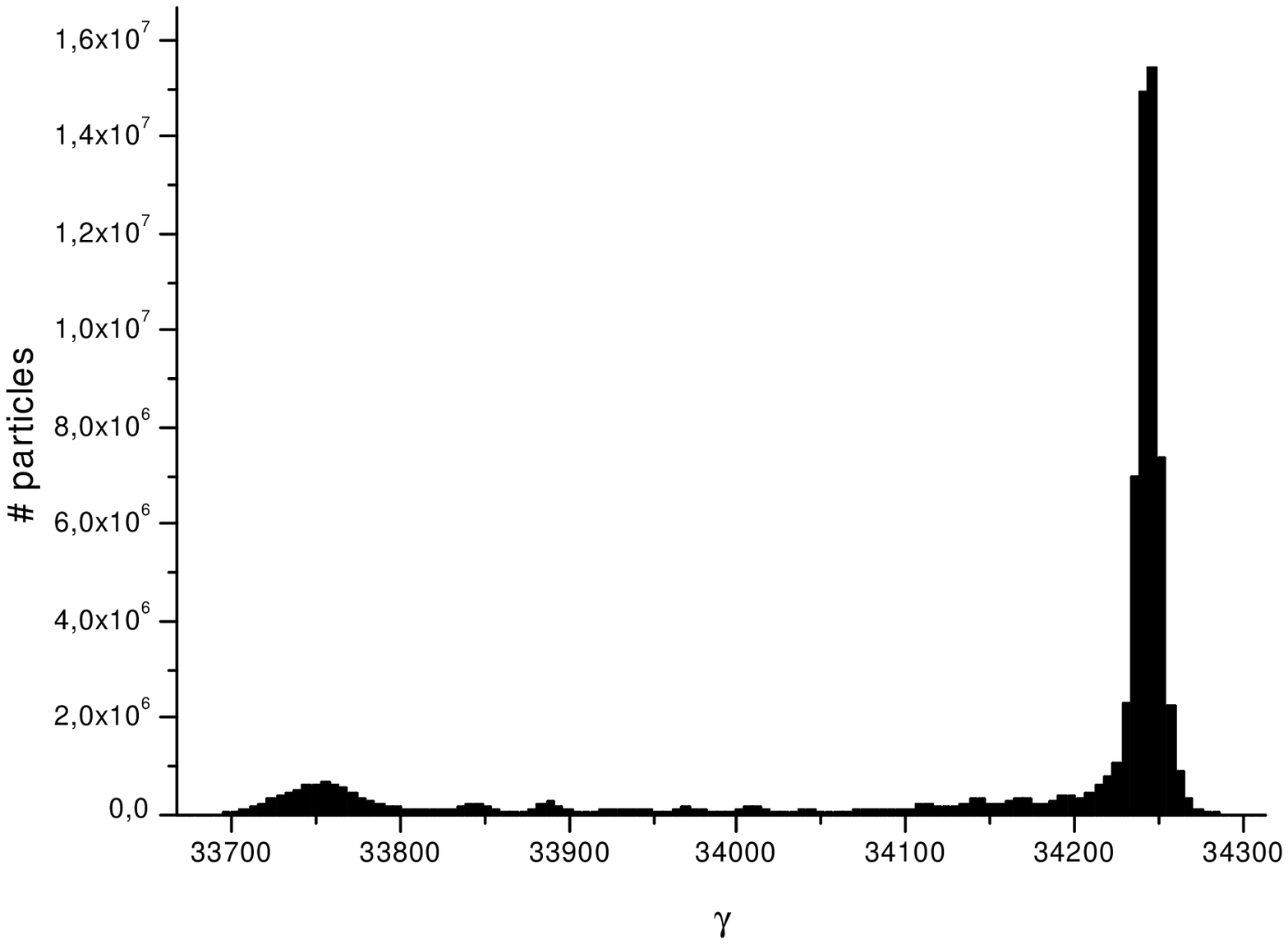}
\caption{Energy distribution of the particles in the
electron bunch at the undulator exit in the case of undulator tapering according to Fig. \ref{Taplaw}.} \label{Distr_total_L}
\end{figure}

Finally, for completeness, it is interesting to have a look at the dynamics of electrons entering the tapering process. As is well known, the tapering process allows for many electrons to be kept into the deceleration region of the longitudinal phase-space, which allows them to contribute to the build-up of the radiation field with their kinetic energy. This is shown in Fig. \ref{Phspace_L}. The black spots represent macroparticles near the electron beam axis. The grey spots all other macroparticles.  The counterparts in terms of energy distributions are shown in Fig. \ref{Distr_middle_L} and Fig. \ref{Distr_total_L}. In particular, Fig. \ref{Distr_total_L} presents the energy distribution in the electron beam at the exit of the undulator, averaged over the electron bunch. The analysis of the longitudinal phase space shown in these figures illustrates how electrons in the central slice of the electron bunch are well separated into two (tapered and untapered) fractions. However, the particles located at the tails of the bunch are not catched in the regime of coherent deceleration. It is seen that the energy spectrum of the spent beam is about $2 \%$.

\section{\label{sec:conc} Conclusions}

This paper discusses the potential to enhance the capabilities of the European XFEL. In the hard X-rays regime, a high longitudinal coherence, in addition to full transverse coherence, will
be the key to a performance upgrade.

The progress which can be achieved with the methods considered here is based on the new self-seeding scheme with wake monochromators proposed in \cite{OURX, OURY2}, which is extremely compact and can be straightforwardly installed in the baseline undulator system of the
European XFEL. The final output radiation bandwidth of the self-seeded XFEL is close to the Fourier limit given by the finite duration of the pulse.

With a radiation beam monochromatized down to the Fourier transform limit, a variety of very different techniques leading to further
improvement of the XFEL performance become feasible. Despite the unprecedented increase in peak power of the X-ray pulses for SASE
X-ray FELs, some applications may require still higher photon flux. The output power of the European XFEL could be straightforwardly increased by tapering the output undulator. Following this line of reasoning, we propose a scheme for generating TW-level,  fully-coherent X-ray pulses in XFELs by exploiting the tunable-gap
baseline SASE2 undulator, and the previously discussed self-seeding scheme. The development of the radiation source described here promises to open up new areas in life sciences by allowing diffraction imaging of any molecule without the need for crystallization.

This paper also describes an efficient way for obtaining a multi-user facility. The high monochromatization of the output radiation constitutes the key for reaching such result. We propose a
photon beam distribution system based on the use of crystals in Bragg reflection geometry as movable deflectors. About $99 \%$ reflectivity can be achieved for monochromatic X-rays. Angular and bandwidth acceptances of crystal deflectors are much wider compared to bandwidth and divergence of the X-ray beam. Therefore, it can be possible to deflect the full radiation pulse of an angle of order of a radian without perturbations. The proposed photon beam distribution system would allow to switch the X-ray beam quickly between many experiments in order to make an efficient use of the source.

\section{Acknowledgements}

We are grateful to Massimo Altarelli, Reinhard Brinkmann, Serguei
Molodtsov and Edgar Weckert for their support and their interest
during the compilation of this work.


\begin{thebibliography}{99}

\bibitem{LCLS1} J. Arthur et al. (Eds.) Linac Coherent Light Source
(LCLS). Conceptual Design Report, SLAC-R593, Stanford (2002) (See
also http://www-ssrl.slac.stanford.edu/lcls/cdr).

\bibitem{LCLS2} P. Emma, First lasing of the LCLS X-ray FEL at 1.5 Å, in
Proceedings of PAC09, Vancouver, to be published in
http://accelconf.web.cern.ch/AccelConf/ (2009).

\bibitem{DING} Y. Ding et al., Phys. Rev. Lett. 102, 254801
(2009).


\bibitem{tdr-2006} M. Altarelli, et al. (Eds.)
XFEL, The European X-ray Free-Electron Laser, Technical Design
Report, DESY 2006-097, Hamburg (2006).


\bibitem{TAP1} A. Lin and J.M. Dawson, Phys. Rev. Lett. 42 2172 (1986)

\bibitem{TAP2} P. Sprangle, C.M. Tang and W.M. Manheimer, Phys. Rev. Lett. 43 1932 (1979)

\bibitem{TAP3} N.M. Kroll, P. Morton and M.N. Rosenbluth, IEEE J. Quantum Electron., QE-17, 1436 (1981)

\bibitem{TAP4} T.J. Orzechovski et al., Phys. Rev. Lett. 57, 2172 (1986)
%

\bibitem{FAWL} W. Fawley et al., NIM A 483 (2002) p 537

\bibitem{OURX} G. Geloni, V. Kocharyan and E.~Saldin, "A simple method for controlling the line width of SASE X-ray FELs",
DESY 10-053 (2010).

\bibitem{OURY2} G. Geloni, V. Kocharyan and E.~Saldin, "A Cascade self-seeding scheme with wake monochromator for narrow-bandwidth X-ray FELs", DESY 10-080 (2010).

\bibitem{GENE} S Reiche et al., Nucl. Instr. and Meth. A 429, 243 (1999).
%
%
%
%
%
%
%
%
%
%
%
%
%
%
%
%
%
%
%





\end{thebibliography}
\end{document}